\documentclass{epl2}

\title{Time variations in the deep underground muon flux}

\author{S.~Cecchini\inst{1,2} \and M.~Cozzi\inst{2,3} \and H.~Dekhissi\inst{4} \and J.~Derkaoui\inst{4} \and G.~Giacomelli\inst{2,3} \and \\
M.~Giorgini\inst{2,3} \and F.~Maaroufi\inst{4} \and G.~Mandrioli\inst{2} \and A.~Margiotta\inst{2,3} \and A.~Moussa\inst{4} \and 
L.~Patrizii\inst{2} \and M.~Sioli\inst{2,3} \and G.~Sirri\inst{2} \and M.~Spurio\inst{2,3} \and V.~Togo\inst{2}}
\shortauthor{S.~Cecchini \etal}

\institute{                    
  \inst{1} IASF/INAF, Sezione di Bologna, I-40129 Bologna, Italy, EU \\
  \inst{2} INFN, Sezione di Bologna, I-40127 Bologna, Italy, EU \\
  \inst{3} Dipartimento di Fisica dell'Universit\`a di Bologna, I-40127 Bologna, Italy, EU \\
  \inst{4} LPTP, Faculty of Sciences, University Mohamed I, B.P.424, Oujda, Morocco
}


\abstract{
More than 35 million high-energy muons collected with the MACRO detector 
at the underground Gran Sasso Laboratory have been used to search for flux 
variations of different nature. Two kinds of studies were carried 
out: a search for the occurrence of clusters of events and a search for periodic variations. 
Different analysis methods, including the Scan Statistics test and the Lomb-Scargle spectral 
analysis have been applied to the data.
}

\begin{document}

\maketitle

\section{Introduction}
\label{introduction}
The high energy muon events collected by the MACRO apparatus at the 
average depth of 3800 m.w.e. represent one of the most extensive records 
of such kind of data. These data can be used to 
search for time variations of periodic and stochastic nature, as 
it was done extensively by using arrival times of Extensive Air Shower \cite{first}.
Variations in the underground muon flux may be due to different causes 
of galactic, solar and terrestrial origin. The common problem for this 
type of searches is to determine whether an observed effect has occurred 
by chance or if it signals a departure from a pure random distribution.

Bursts of underground muon events may be originated by violent events as GRB on the 
time scale of few seconds or by sudden metereological variations on the scale of 
several hours. Periodic modulations are related to the cooling of
the upper atmosphere during the night (winter) producing daily (seasonal) variations
of the atmospheric density and therefore in the underground muon flux.
A sidereal modulation may be introduced by the Earth roto-traslation in a
isotropical cosmic ray distribution (Compton-Getting effect).

MACRO was a multipurpose modular apparatus with 6 supermodules equipped
with liquid scintillators, limited streamer tubes and nuclear track 
detectors \cite{MACRO}. The complete detector was a nearly ``closed box'' with a
total length of 76.7 m; it had a lower part 4.8 m high, with 10 horizontal layers
of streamer tubes, two layers of scintillators and seven layers of rock absorber.
The upper part (``attico'') was 4.5 m high and contained 4 layers of streamer tubes,
one layer of scintillators and the electronics. The streamer tube system was used
for tracking, and provided two independent views: the wire view and the strip view.
The latter employed 3 cm wide aluminium strips at 26.5$^{\circ}$ with respect to
the wire view.

MACRO studied atmospheric neutrinos and their oscillations \cite{neu,lorentz}, 
various aspects of cosmic ray physics and astrophysics \cite{crmu}, 
searched for GUT Magnetic Monopoles \cite{MM} and other exotica \cite{exotica}.
Some interruptions of different kinds occurred during data taking, either 
randomly (e.g. power outages), or regularly (e.g. maintenance), so 
appropriate statistical methods have to be applied and particular 
care should be used in choosing periods of stationary conditions.

In the following we discuss the results of the searches for periodic 
variations and for time clustering of muon events.

\section{Burst search}
\label{burst}

Using the MACRO data we performed a new analysis on muon time series in order to find isolated clusters of events.
The MACRO Collaboration published the results of a previous analysis obtained with
another method, namely the study of the time interval distribution \cite{erlangen}.
For each muon arriving at time t${_0}$ the distribution of the 
time interval elapsed between the first muon t${_0}$ and the next five muons: 
t${_i}$-t${_0}$, i=1,...,5 was calculated.

\begin{figure}[t]
\onefigure[scale=0.4]{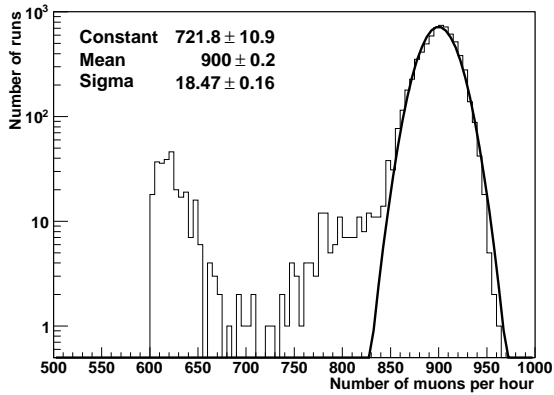}
\caption{Distribution of the muon rate per run, in units of $\mu$/hour.
We accepted runs within 3$\sigma$ from the average of the gaussian fit superimposed to the data.
The small peak on the left corresponds to	runs with the detector partially in acquisition
(1/3 of the supermodules were off). The tail at the left of the primary peak ($\sim$1\% of the total) corresponds to 
runs where one module was in maintenance and the data taking was still progressing for particle searches.}
\label{fig:fig0}
\end{figure}

For the present analysis we considered data recorded by the streamer tube 
system in the time interval November 1991-May 2000 and we selected the 
data with the following criteria:

- run duration longer than 1 hour;

- streamer tube efficiency larger than 90\% for the wire view
and 85\% for the strip view. The streamer tube efficiency was obtained
using the sub-sample of muon tracks crossing all the 10 lower streamer tube planes;

- all 6 super-modules in acquisition.

- acquisition dead time smaller than 2.5\% for the whole detector.

- runs having muon counting rates that deviated more than 3$\sigma$ from the average
were removed (Fig. \ref{fig:fig0}). It was checked that this cut did not bias the 
analysis: for the Scan Statistics, the expected burst durations are small compared
to the run durations; for the periodicity search, the expected modulation amplitudes are 
negligible compared to the fluctuations around the average number of muons per hour.

- we required a single reconstructed track in each one of the two projective views.
This cut excludes muon bundles from the analysis and provides a clean sample of single muon events.

The total number of runs surviving the cuts was 6113 corresponding
to $3.5 \cdot 10^7$ muon events. This sample was used to search for bursts
of cosmic ray particles coming from the whole upper hemisphere.
In this analysis, we looked for bursts of events using the so-called ``scan statistics''
method. This is a bin-free method and it provides unbiased results
(see \cite{scan} and references therein).

Let us consider an interval $\Delta \equiv [A,B]$ of a continuous variable $x$ 
and a Poisson process with density $\lambda$.
We call {\it scan statistics} (SS) the largest number of events found in any 
subinterval of $[A,B]$ of length $\omega$
\begin{equation}
S(w) \equiv \max_{{\mathcal{A}}\leq x \leq {\mathcal{B}}-\omega} \left\{ Y_x(w) \right\} 
\end{equation}
where $Y_x(w)$ denotes the number of events in the subinterval $[x,x+w]$.
The probability $P$ that a statistical fluctuation would produce a burst 
of events as large as $k$ can be approximated by \cite{scan}:
\begin{equation}
P = 1 - Q^{*}(k;\psi L,1/L) \simeq 1 - Q^{*}_2\left[Q^{*}_3/Q^{*}_2\right]^{L-2}
\label{equ:naus}
\end{equation}
where $\psi\equiv \lambda w$, $L=\Delta/w$ and
\begin{eqnarray}
Q^{*}_2 & = & \left[F(k-1,\psi)\right]^{2}-
\left(k-1\right)p(k,\psi)p(k-2,\psi) \nonumber \\
& & -\left(k-1-\psi\right)p(k,\psi)F(k-3,\psi)
\label{equ:Q2_naus} \\
Q^{*}_3 & = & \left[F(k-1,\psi)\right]^{3}-A_{1}+A_{2}+A_{3}-A_{4} 
\label{equ:Q3_naus} 
\end{eqnarray}
with
\begin{eqnarray*}
A_{1} & = & 2 p(k,\psi)F(k-1,\psi)\{\left(k-1\right)F(k-2,\psi) \nonumber \\ 
& & -\psi F(k-3,\psi)\} \\
A_{2} & = & 0.5 \left[ p(k,\psi) \right]^{2} \{ (k-1)(k-2)F(k-3,\psi) \nonumber \\ 
A_{3} & = & \sum_{r=1}^{k-1}p(2k-r,\psi)\left[F(r-1,\psi)\right]^{2} \nonumber \\
A_{4} & = & \sum_{r=2}^{k-1}p(2k-r,\psi)p(r,\psi)\{(r-1)F(r-2,\psi) \nonumber \\
& & -\psi F(r-3,\psi)\}
\end{eqnarray*}
In the above formulas $F(k,\psi)$ denotes the cumulative distribution
\begin{equation}
F(k,\psi) = \sum_{i=0}^k p(i,\psi) \ \ ; \ \ p(i,\psi)=e^{-\psi} 
\frac{\psi^i}{i!}
\end{equation}
and $ F(k,\psi)=0$ for $k<0$.

\begin{figure}[t]
\onefigure[scale=0.4]{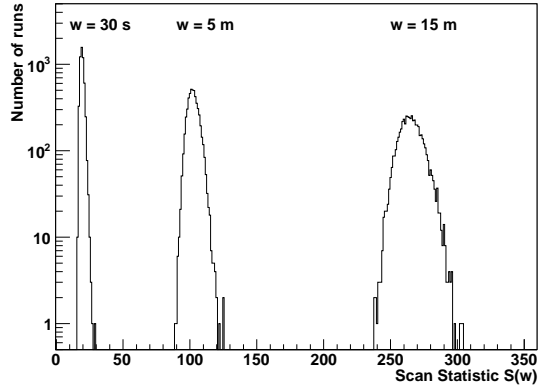}
\caption{Distribution of the $S(w)$ SS variable for the three sliding windows used in this work.
Each entry of the histograms corresponds in the horizonthal scale to the maximum number of events falling within a given time window
for each run.}
\label{fig:fig1}
\end{figure}

We used the SS in the following way: for each 
run $i$, let $[A_i,B_i]$ be the time interval ranging from the 
start to the end of the run. The interval was scanned counting the number of muons 
inside a ``time window'' $w$ of fixed length. Let $k_i$ be 
the maximum number of events recorded during the scan. For each run, 
according to Eq. \ref{equ:naus}, we computed the probability $P_i$ that a statistical fluctuation 
would produce a burst of events as large as $k_i$. The only parameter to be fixed ``a priori'' 
is the scanning window $w$, whose size must be chosen on the basis of astrophysical considerations
(e.g. time duration of typical HE burst events).
We tried different sizes ($w=30$ s, 5 min and 15 min) and for each the probability 
distribution $P_i$ (i=1, N$_{run}$) was analyzed.

Fig. \ref{fig:fig1} shows the $S(w)$ distributions for the three different values of the scanning
window $w$.

\begin{figure}
\onefigure[scale=0.4]{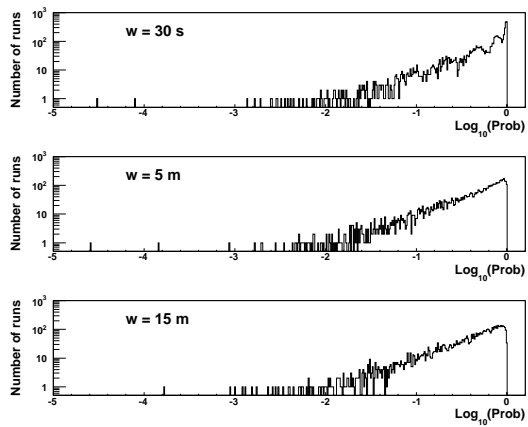}
\caption{SS probability distributions for the selected runs. 
In the upper panel a time window $w=30$ s was used; $w=5$ min in the central 
panel and  $w=15$ min in the bottom panel.}
\label{fig:fig2}
\end{figure}

In Fig. \ref{fig:fig2} we show the probability distribution for the 6113 runs surviving the analysis cuts:
$w=30$ s (Fig. \ref{fig:fig2} top), 5 min (Fig. \ref{fig:fig2} middle) and 15 min (Fig. \ref{fig:fig2} bottom).
No significant deviations from the null hypothesis was found.
The analyses of unusual runs with probabilities smaller than $5\cdot 10^{-4}$ have shown that ``bumps of events''
were located near the beginning or the end of the runs. The structures at Log$_{10}$(Prob) $>-0.4$
in the plot with $w$=30 s are related to the time window's width comparable with the muon
rate of $\sim$0.2 muons/s. In this case non-Poissonian sources due to background,
dead time, etc. start to dominate and cause a deviation from the expected distribution.

\section{Periodicity search and spectrum analyses}
\label{lomb}

\begin{figure}
\onefigure[scale=0.4]{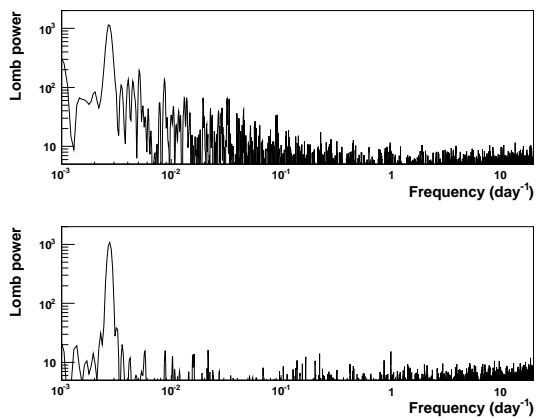}
\caption {Lomb power $p$ as a function of the frequency
[days$^{-1}$] for experimental data (upper panel). Note the high peak 
at $\sim$2.7$\cdot 10^{-3}$ corresponding to 365 days (seasonal flux variation). In 
the lower panel the results of a Monte Carlo simulation having the same 
noise level of the real data with seasonal, solar diurnal and sidereal 
waves \cite{seasonal,sidereal} added.}
\label{fig:fig3}
\end{figure}

The Fourier amplitude spectrum analysis is a powerful technique that allows 
a blind search for regular/persistent fluctuations in time 
series \cite{attolini}. Such a technique, however, requires the input data 
to be sampled at evenly spaced intervals; data gaps of variable length 
and occurring randomly in the series produce spurious contributions to 
the power that can mimic the presence of a periodicity.

The Lomb-Scargle method \cite{lomb} mitigates this effect, even in the 
case of very long data series. Moreover, as indicated in Ref. \cite{prob}, it 
allows to evaluate the significance of the ``peaks'' (signal) 
with respect to the null hypothesis.
The data used in this analysis were collected with the full detector and with strict 
selection criteria.
The muon events were binned in 15 min time intervals and bins deviating 
by more than $3 \sigma$ from the monthly average rate were discarded. The 
total number of time-bins used was 160242 corresponding to 58\% of the 
whole sample.

\begin{figure}
\onefigure[scale=0.4]{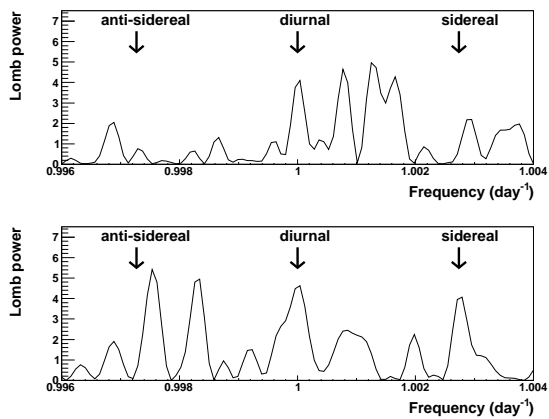}
\caption{The frequency region around the solar diurnal wave. The arrows 
mark its position and also the sidereal and anti-sidereal peaks. The upper panel
refers to real data, the lower panel to Monte Carlo simulation.}
\label{fig:fig4}
\end{figure}

The results of our analysis are shown in Fig. \ref{fig:fig3}. We compare 
the spectrum obtained for the real data (upper panel) with a Monte Carlo simulation having 
the same noise level and time intervals distributed according to the sequence 
of the original series (lower panel).
Seasonal, solar diurnal and sidereal waves were added in the simulated series with 
the amplitudes found in Refs. \cite{seasonal,sidereal}.
Real data have a power distribution which decreases with frequency$^{-1/2}$
up to $f$=1 day$^{-1}$: this behavior, mainly related to the dead time of the
apparatus, was not included in the simulation.
The most striking feature of the spectrum (note the logarithmic vertical scale) 
is the large peak at $\sim2.7 \cdot 10^{-3}$ 
corresponding to the seasonal variation of the muon flux as found in Refs. \cite{seasonal,sidereal}.

Fig. \ref{fig:fig4} shows a frequency region around the solar diurnal frequency 
where are indicated the frequencies corresponding to the sidereal 
and anti-sidereal waves, both for data (upper panel) and Monte Carlo simulation
(lower panel). In order to ``clean'' the power spectrum from
unphysical frequencies, we performed a ``sliding average'' both for data 
and Monte Carlo: each time-bin was rescaled according to the formula:
\begin{equation}
N^{'}_{i} = \frac{N_{i} - \bar{N}(\Delta \tau)}{\bar{N}(\Delta \tau)}
\label{sliding}
\end{equation}
where $N_i$ is the original bin content and $\bar{N}(\Delta \tau)$ is the average
content in the time range $\pm \Delta \tau$ (we chose $\Delta \tau$ = 1 day).
The peak at frequency 1 day$^{-1}$ has a statistical 
significance of $\sim$2.3 $\sigma$; the statistical significance assuming
an oscillatory behavior is  $\sim$3.4~$\sigma$.
We looked also in the frequency regions around the sidereal 
and anti-sidereal waves. A signal corresponding to the sidereal variation
is observed, but peaks of similar size (or even larger) are present elsewhere in the spectrum. 
The claim that the sidereal and solar diurnal waves are real is based on their occurrence 
at a frequency of ``a priori'' interest and on the stability of its 
amplitude and phase with time: this is an indication of an independent
observation of the same modulations found with the ``folding'' method described in
Ref. \cite{sidereal}. The amplitudes and the probabilities 
for the null hypothesis computed with the folding method and with the
Lomb-Scargle method are in fair agreement.

\section{Conclusions}
\label{conclusions}
We analyzed the time series of MACRO muons using two complementary 
approaches: search for bursts of muon events and search for periodicities in the muon time distribution.
The SS method was used in the first case and  the Lomb-Scargle method in the 
second case. The two techniques complete early analyses performed with ``folding''
methods in searching for periodicities and time differences for burst 
events. The seasonal modulation (Fig. \ref{fig:fig3}) was confirmed. 
The signals at the positions of the solar diurnal and sidereal modulations are 
confirmed, even if with smaller significance (Fig. \ref{fig:fig4}).
No other deviations from the expected distributions were found.

\acknowledgments
We thank the members of the MACRO Collaboration and the personnel of the LNGS 
for their cooperation. We acknowledge the collaboration between the Universities of 
Bologna and Mohamed I of Oujda. We thank INFN for providing FAI fellowships.

\end{document}